%% file: main.tex
\documentclass[conference,a4paper]{IEEEtran}

\PassOptionsToPackage{nolist}{acronym}
\usepackage{acronym}
\usepackage{amsmath,amssymb,amsfonts}
\usepackage{graphicx}
\usepackage{textcomp}
\usepackage{xcolor}
\usepackage{tikz}
\usepackage{pgfplots}
\pgfplotsset{compat=1.18}
\usetikzlibrary{arrows, arrows.meta, calc, positioning, quotes, shapes}
\usepackage[position=bottom]{subfig}
\usepackage[noadjust]{cite}
\UseRawInputEncoding

\usepackage{amsmath}
\usepackage{algorithm}
\usepackage[noend]{algpseudocode}
\algrenewcommand\algorithmicindent{0.5em}

\usepackage{xcolor,colortbl}
\begin{document}


\begin{acronym}[ECU]
        \acro{AQM}{Active Queue Management}
        \acro{RED}{Random Early Detection}
        \acro{CoDel}{Controlled Delay}
        \acro{RTT}{Round Trip Time}
        \acroplural{RTT}[RTTs]{Round Trip Times} 
        \acro{sRTT}{Smoothed Round Trip Time}
        \acroplural{sRTT}[sRTTs]{Smoothed Round Trip Times}
        \acro{NR}{New Radio}
        \acro{NSA-NR}{Non-Standalone New Radio}
        \acro{LTE}{Long Term Evolution}
        \acro{PTP}{Precision Time Protocol}
        \acro{HARQ}{Hybrid Automatic Repeat Request}
        \acro{SCS}{Subcarrier Spacing}
        \acro{DRX}{Discontinuous Reception Mode}
        \acro{NSA}{non-standalone}
        \acro{RRC}{Radio Resource Control}
        \acro{ACK}{Acknowledgement}
        \acroplural{ACK}[ACKs]{Acknowledgements}
        \acro{NACK}{Not Acknowledged}
        \acro{ARQ}{Automatic Repeat reQuest}
        \acro{recvwnd}{Receive Window}
        \acro{UE}{User Equipment}
        \acroplural{UE}[UEs]{User Equipments}
        \acro{cwnd}{congestion window}
        \acroplural{cwnd}[cwnds]{congestion windows}
        \acro{ICMP}{Internet Control Message Protocol}
        \acro{CBR}{Constant Bit Rate}
        \acro{NAT}{Network Address Translation}
        \acro{UDP}{User Datagram Protocol}
        \acro{OWD}{One Way Delay}
        \acroplural{OWD}[OWDs]{One Way Delays}
        \acro{BBR}{Bottleneck Bandwidth and Round-trip time}
        \acro{TCP}{Transmission Control Protocol}
        \acro{CDF}{Cumulative Distribution Function}
        \acro{eMBB}{Enhanced mobile broadband}
        \acro{mMTC}{Massive machine-type communication}
        \acro{URLLC}{Ultra-reliable and low-latency communication}
        \acro{TB}{Transport Block}
        \acro{FEC}{Forward Error Correction}
        \acro{CBG}{Codeblock Group}
        \acroplural{CBG}[CBGs]{Codeblock Groups}
        \acro{OFDM}{Orthogonal Frequency Division Multiplexing}
        \acro{SCS}{Subcarrier Spacing}
        \acro{SCC}{Secondary Component Carrier}
        \acroplural{SCC}[SCCs]{secondary component carriers}
        \acro{CC}{congestion control}
        \acro{RSRQ}{Reference Signal Received Quality}
        \acro{RAN}{Radio Access Network}
        \acro{SINR}{Signal-to-Interference-plus-Noise Ratio}
        \acro{ECN}{Explicit Congestion Notification}
        \acro{QUIC}{Quick UDP Internet Connections}
        \acro{BDP}{Bandwidth-Delay Product}
        \acroplural{BDP}[BDPs]{Bandwidth-Delay Products}
        \acro{RLC}{Radio Link Control}
        \acro{ROCCET}{RTT oriented CUBIC congestion control exTension}
        \acro{srRTT}{smoothed relative RTT}
        \acro{rRTT}{relative RTT}
        \acro{EWMA}{Exponentially Weighted Moving Average}
        \acro{BIC}{Binary Increase Congestion control}
        \acro{ORBITER}{Optimal Rate Balancing with Incremental Throughput Enhancement and Reliability}
        \acro{LAUNCH}{Latency-Aware Utilization for Network Congestion Handling}
        \acro{CE}{congestion event}
        \acroplural{ce}[ces]{congestion events}
        \acro{SA}{Stand Alone}
\end{acronym}

\acused{TCP}
\acused{RTT}
\acused{ACK}
\acused{LTE}
\acused{NR}

\title{TCP ROCCET: An RTT-Oriented CUBIC Congestion Control Extension for 5G and Beyond Networks}
\author{\IEEEauthorblockN{Lukas Prause and Mark Akselrod}
\IEEEauthorblockA{\textit{Institute of Communications Technology} \\
\textit{Leibniz Universit\"at Hannover}\\ 
\{lukas.prause, mark.akselrod\}@ikt.uni-hannover.de
}

}

\maketitle

\begin{abstract}
The behavior of loss-based TCP congestion control algorithms like TCP CUBIC continues to be a challenge in modern cellular networks. Due to the large RLC layer buffers required to deal with short-term changes in channel capacity, the behavior of both the Slow Start and congestion avoidance phases may be heavily impacted by the lack of packet losses and the resulting bufferbloat. While existing congestion control algorithms like TCP BBR do tend to perform better even in the presence of large bottleneck buffers, they still tend to fill the buffer more than necessary and can have fairness issues when compared to loss-based algorithms.

In this paper, we analyze the issues with the use of loss-based congestion control algorithms by analyzing TCP CUBIC, which is currently the most popular variant. To mitigate the issues experienced by TCP CUBIC in cellular networks, we introduce TCP ROCCET, a latency-based extension of TCP CUBIC  that responds to network congestion based on round-trip time in addition to packet loss. 

Our findings show that TCP ROCCET can reduce latency and bufferbloat compared to the standard CUBIC implementation, without requiring a specific network architecture. Compared to TCP BBRv3, ROCCET offers similar throughput while maintaining lower overall latency. The evaluation was conducted in real 5G networks, including both stationary and mobile scenarios, confirming ROCCET's improved response to network congestion under varying conditions.
\end{abstract}

\section{Introduction}

\noindent As the maximum data rates achievable in current \acp{RAN} continue to grow, the size of the \ac{RLC} layer buffers that is needed to deal with bursty TCP traffic and short-term signal quality variations also increases. While the larger \ac{RLC} layer buffer size is required when dealing with the traffic when the signal quality is at its highest, long-term decreases can lead to a standing queue at the \ac{RLC} layer buffer, resulting in bufferbloat. 

In cases where the available bandwidth or the capacity of a link decreases, loss-based congestion control algorithms like TCP CUBIC \cite{cubic} expect the bottleneck buffer to start dropping packets as their sending rate exceeds the rate at which the packets can be processed. However, the deep \ac{RLC} layer buffer tends to just be filled up by the initially excessive sending rate, and then the sender self-adjusts their sending rate to the ACK-rate. In the case that the initial excessive sending rate is not enough to cause a packet loss, TCP's self-clocking prevents the buffer from ever being drained of the excessive packets.

This problem can be particularly severe in current Non-Standalone-New Radio (NSA-NR) networks: In such deployments, a \ac{UE} is typically connected to multiple carriers at the same time. As the \ac{UE} changes its position, it can move out of range of one of the connected carriers, thus losing all of its bandwidth. This can lead to severe changes in the channel capacity, especially if the connection to the \ac{NR}carrier is lost. We have previously explored this problem in a major commercial NSA-NR network by conducting a mobile and stationary measurement campaign~\cite{prause_2023}. However, this problem is not limited to NSA-NR networks, since LTE is expected to continue to be used as a fallback for NR, and even in NR-only scenarios, multiple \ac{NR} carriers can also be connected using carrier aggregation.

Many different ways have been proposed to deal with this problem. On the sender's side, novel congestion control algorithms like TCP \ac{BBR} \cite{bbr} or Copa \cite{arun2018copa} adjust their sending rate in a way that tries to avoid filling up the buffer. \ac{AQM} algorithms like RED \cite{251892} or CoDel \cite{RFC8289} can be deployed at the \ac{RLC} layer buffer to try to drop or mark packets if the buffer is too full or the sojourn time is too high. Finally, some smartphone manufacturers limit the receive window when a cellular connection is used, thus limiting the maximum size of the sender's send window and the amount of bufferbloat.

In this paper, we propose a delay-based extension to the popular loss-based CUBIC algorithm called \ac{ROCCET}. In contrast to using an \ac{AQM} method, a sender-based solution does not require the service provider to rely on a particular network architecture to reduce delays. We evaluate \ac{ROCCET} in both stationary and mobile scenarios. We compare it to both the base CUBIC implementation and TCP BBRv3 \cite{bbrv3} as these two are some of the most popular congestion control algorithms currently in deployment. Our measurements show that TCP ROCCET can achieve significantly lower \acp{RTT} than CUBIC while maintaining a high throughput. When compared to TCP BBRv3, ROCCET also achieves a similar throughput while keeping the latency lower.

The rest of the paper is organized as follows. In Section \ref{sec:primer-cubic}, we provide a short primer on the functionality of TCP CUBIC. In Section \ref{sec:cc-issues}, we discuss the issues that we have observed when using TCP CUBIC in mobile cellular networks. In Section \ref{sec:roccet-algo}, we present \ac{ROCCET}, our delay-based extension for TCP CUBIC, to mitigate these issues. In Section \ref{sec:measurement_setup}, we present our real-world measurement setup that we use to evaluate \ac{ROCCET}'s performance in cellular networks in stationary and mobile scenarios. We present and discuss our stationary and mobile measurement results in Sections \ref{sec:stat_measurements} and \ref{sec:mob_measurements}. In Section \ref{sec:fairness}, we evaluate TCP ROCCET's bandwidth sharing behavior with CUBIC and BBRv3. In Section \ref{sec:bandwidth_change_reaction}, we present a detailed view of how TCP ROCCET behaves when the capacity of a link changes when compared to TCP CUBIC and BBRv3. We discuss the limitations of our algorithm and considerations for alternative parameter choices to tune its behavior in Section \ref{sec:discussion}. Section \ref{sec:fairness} evaluates \ac{ROCCET}'s fairness compared to itself, TCP CUBIC, and TCP BBRv3. Finally, Section \ref{sec:conclusion} concludes the paper and provides an outlook on further development and rollout steps.

\section{A Short TCP CUBIC Primer}
\label{sec:primer-cubic}
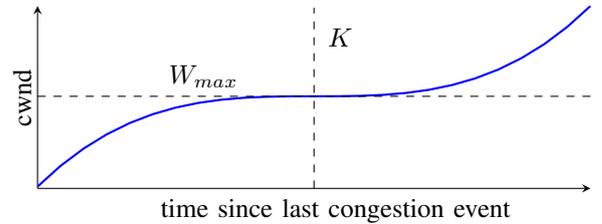
\begin{figure}
    \centering
    \begin{tikzpicture}
    \begin{axis}[
        width=\linewidth,height=4cm,
        xlabel={},
        ylabel={},
        domain=-10:10,
        axis lines=left,
        xtick=\empty,
        ytick=\empty,
    ]
        \addplot [
            color=blue,
            thick
        ]
        {(x)^3 + 20}; 

        \addplot [
            color=black,
            dashed
        ]
        {20} node[above,pos=0.3] {$W_{max}$};
        \addplot[color=black, dashed] coordinates {(0,-1000)(0,1000)};
    \end{axis}
    \draw [fill] (3.7,2) node [right] {$K$};
    \draw [fill] (1.5,-0.3) node [right] {time since last congestion event};
    \draw [fill] (-0.2,.7) node [right,rotate=90] {cwnd};
\end{tikzpicture}
    \caption{Characteristic \ac{cwnd} growth function of TCP CUBIC. With the \ac{cwnd} size before the last congestion event ($W_{max}$) as a saddle point. The saddle point is reached at time $K$ after the last congestion event.}
    \label{fig:cubic-window-growth}
\end{figure}

\noindent In this section, we provide a short primer on TCP CUBIC and its functionality.
TCP CUBIC is an extension of TCP \ac{BIC} \cite{tcp-bic}. \ac{BIC} was introduced to better utilize high-speed networks with large \acp{BDP}.
Without going into detail, BIC uses binary search to find a \ac{cwnd} for the available bandwidth.
For that, the \ac{cwnd} before a \ac{CE} $W_{max}$ and the \ac{cwnd} after a reduction $W_{min}$ are considered. If the calculated increase is larger than a predefined maximum increment per RTT $S_{max}$, $S_{max}$ gets added to the \ac{cwnd} size instead of the calculated increment.
If the \ac{cwnd} size exceeds $W_{max}$, BIC enters a "max probing" phase. In this phase, BIC uses a \ac{cwnd} growth function which is symmetric to the growth function before.
A downside of this \ac{cwnd} calculation is that it is too aggressive for short RTT or low-speed networks.
At this point, TCP CUBIC provides a solution for these problems by approximating the \ac{cwnd} growth function of BIC with a cubic function. TCP CUBIC adopts the value of $W_{max}$ and uses it as the saddle point for the cubic \ac{cwnd} growth function. This results in the cwnd being adjusted more cautiously when it approaches the point of the last \ac{CE}, and increasingly more aggressively the further away the \ac{cwnd} is from $W_{max}$.
In addition, TCP CUBIC's \ac{cwnd} growth function is independent of the RTT, because only the time since the last \ac{CE} is considered.
The characteristic \ac{cwnd} growth function of TCP CUBIC is shown in Fig.~\ref{fig:cubic-window-growth} and the corresponding equation is shown in Eq.~(\ref{eq:cubic-window}). The parameters $\beta$ and $C$ are implementation dependent, $\beta$ is the reduction factor of the \ac{cwnd} after a \ac{CE}, and $C$ is a scaling factor of the \ac{cwnd}.
In this paper, we refer to the TCP CUBIC implementation of the Linux kernel, where TCP CUBIC additionally uses HyStart~\cite{tcp-hystart}, an extension of Slow Start. HyStart differs from Slow Start by not only relying on the TCP Slow Start threshold as the beginning of the congestion avoidance phase. HyStart additionally monitors the inter-packet gap of \acp{ACK} and the \ac{RTT} to find an early beginning of congestion, to start into the congestion avoidance phase.

\begin{equation}
\begin{aligned}
W(t) &= C \cdot (t - K)^3 + W_{max}
\\
K &= \sqrt[3]{\frac{W_{max} \cdot \beta}{C}}
\label{eq:cubic-window}
\end{aligned}
\end{equation}



\section{Known Congestion Control Issues in 4G and 5G Cellular Networks}
\label{sec:cc-issues}
\noindent In this section, we summarize congestion control issues caused by deep RLC layer buffers and carrier aggregation in an \ac{NSA}-NR network, which we have analyzed in detail in our previous work~\cite{prause_2023}. Furthermore, we highlight problems in the Linux kernel implementation of CUBIC.
In our previous measurement analysis, none of the evaluated congestion
control variants, i.e., Reno, CUBIC, and BBRv1, were able to reliably react to changes in carrier bandwidth. Especially the deactivation of the wide 5G carrier through carrier aggregation usually did not result in a large enough reduction of the congestion window, which led to a major increase in delays due to the resulting bufferbloat. Similarly, for Reno and CUBIC, an activation of the 5G carrier did not result in a faster growth of the congestion window, leading to wasted link capacity. Small changes in carrier bandwidth also did not always cause an adjustment of the congestion window, resulting either in bufferbloat or unused link capacity.

For the Linux Kernel implementation of CUBIC, we have observed additional issues. CUBIC can use HyStart or Slow Start, and for both, we have observed several problems:\\

\noindent \textbf{HyStart}: is an extension of Slow Start that additionally observes spacings between ACKs to find a good start for congestion avoidance. Because of the nature of cellular networks (shared medium, scheduled, and wireless), inter-packet gaps are not sufficiently meaningful for congestion detection, which means that the use of HyStart most often leads to a premature start of congestion avoidance, as has already been observed~\cite{hystart1, Akselrod_2020, prause_2022}. This premature start of congestion avoidance leads to an underutilization of available bandwidth. Therefore, Slow Start is a better option for transmissions over 5G networks, which leads to the next issue.

\noindent \textbf{Slow Start}: doubles the \ac{cwnd} during each \ac{RTT} until a threshold is reached or a \ac{CE} in the form of loss occurs.
In the Linux kernel implementation of CUBIC, the initial threshold is the maximum value for an unsigned 32-bit long integer, which in this scenario would be equivalent to infinity. The Slow Start threshold is initially set to this value to prevent too early a start of the congestion avoidance phase.
Therefore, a loss has to occur; otherwise, CUBIC won't enter congestion avoidance. Since cellular networks use deep buffers, Slow Start has to fill a buffer that is possibly too large for the connection before a packet loss can occur. If CUBIC can fill the buffer, the resulting loss will often only partially drain the buffer, which can lead to bufferbloat. 

\begin{figure}
	\centerline{
			\includegraphics[scale=.45]{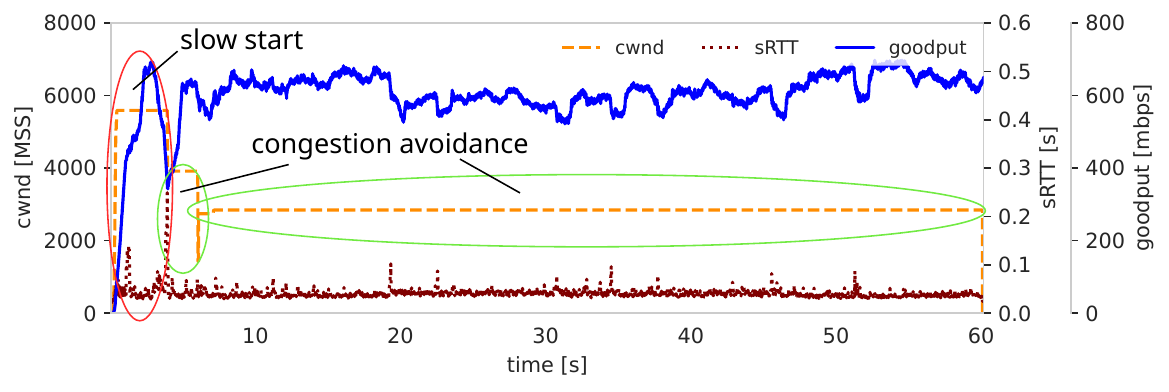}
	}
	\caption{TCP CUBIC first gets stuck in Slow Start and then in congestion avoidance during an unwanted side effect of the Linux kernel implementation. The two loss-based \acs{CE} in the first seven seconds of the transmission are caused by L2 mechanisms and are not related to congestion.}
	\label{fig:cubic_slow-start_example}
\end{figure}

Another implementation detail of CUBIC that can lead to problems during Slow Start is that before CUBIC increases the \ac{cwnd}, it checks if the application layer generates enough data for the current \ac{cwnd}. If this isn't the case, CUBIC does not increase the \ac{cwnd}. In other words, the \ac{cwnd} is frozen until enough data is generated to utilize the current \ac{cwnd}. This often leads to CUBIC never exiting Slow Start and the \ac{cwnd} being stuck on a value that is too large for the current connection.
In some cases, a loss may occur, but this loss is mostly caused by the L2 mechanism DRX or HARQ, which triggers retransmissions by inter-arrival time jitter. In other words, this loss is not related to congestion.
Nevertheless, CUBIC then enters the congestion avoidance phase. In this state, due to the absence of loss, the cwnd is still frozen at a too high value. An exemplary real-world measurement showing the described behavior can be seen in Fig. \ref{fig:cubic_slow-start_example}. We choose an example transmission where loss in Slow Start occurs, to additionally show the lock \ac{cwnd} behavior in congestion avoidance. 

The sample transmission is 60\,s long. In the first 4\,s of the transmission, CUBIC is in Slow Start, which massively overshoots a desirable \ac{cwnd}. Then a loss occurs, and CUBIC enters the first congestion avoidance phase. At 7\,s the next loss occurs, and after a fast retransmit and fast convergence, CUBIC stays in the congestion avoidance phase. During the one-minute-long transmission, except for the first few \acp{RTT} of Slow Start, the \ac{cwnd} is frozen at a value that is too high.

\section{TCP ROCCET}
\label{sec:roccet-algo}
\noindent Taking into account the issues we mentioned in Sec.~\ref{sec:cc-issues}, we developed an RTT-oriented CUBIC congestion control extension (ROCCET).
ROCCET relies on the change of \ac{sRTT}, which is an internal kernel calculation of the \ac{ACK} \ac{RTT} smoothed by an \ac{EWMA}, and the amount of \acp{ACK} arriving during a time period.
With our implementation of ROCCET, we solve the locked cwnd of CUBIC in Slow Start and congestion avoidance. We extend the current Linux kernel implementation of CUBIC by adding two additional metrics for detecting congestion on the network.
The first metric is the \ac{srRTT}. For this, ROCCET calculates the relative change of the \ac{sRTT} ($RTT_{curr}$) compared to the measured minimum sRTT ($RTT_{min}$). After this calculation, the result gets smoothed using an \ac{EWMA} with $\alpha$ as weight factor. This smoothing results in the \ac{srRTT}. The calculation is shown in Eq.~(\ref{eq:relative-change}).
For the required $RTT_{min}$, \ac{ROCCET} compares each ACK RTT with the current $RTT_{min}$. If the current RTT is lower than $RTT_{min}$, it gets updated.

The second metric is the amount of \acp{ACK} that were received over a time interval, compared to the cumulative sum of the last \acp{cwnd}. We further call this sum of \acp{cwnd} \textit{cum\_cwnd}.
The \textit{cum\_cwnd} is calculated over the same time interval as the monitoring of the amount of \acp{ACK}.
The length of the time interval depends on the current phase.

\begin{equation}
\begin{aligned}
x_t &=  \frac{RTT_{curr} - RTT_{min}}{RTT_{min}}
\\
srRTT_t &= \alpha \cdot x_t + (1 - \alpha) \cdot srRTT_{t-1}
\label{eq:relative-change}
\end{aligned}
\end{equation}

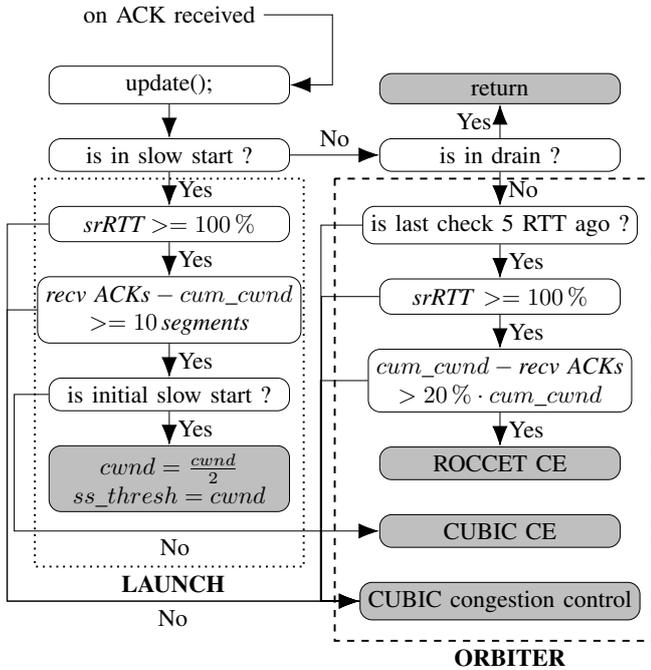
\begin{figure}
    \centering
    \resizebox{\linewidth}{!}{
        \input{figure/roccet_flow_chart}
    }
    \caption{Schematic flow chart of the \acs*{ROCCET} algorithm. Split in the Slow Start phase extension \acs*{LAUNCH} and the congestion control phase \acs*{ORBITER}. Gray states are end states.}
    \label{fig:roccet-flow-chart}
\end{figure}

\noindent Using these metrics, ROCCET can be partitioned into the phases: \ac{LAUNCH}, which is an extension of Slow Start, and the congestion avoidance phase \ac{ORBITER}. A diagram of the mechanism is shown in Fig.~\ref{fig:roccet-flow-chart}.\\

\noindent \textbf{\ac{LAUNCH}} For the exit out of the Slow Start, \ac{ROCCET} tracks the amount of incoming \acp{ACK} and the cum\_cwnd over 100\,ms intervals. If the difference of the incoming \acp{ACK} and the cum\_cwnd is larger than ten segments, and in addition the srRTT is more than 100\,\%, \ac{ROCCET} will exit Slow Start.
If it is the initial Slow Start of the transmission, the \ac{cwnd} gets halved; otherwise, a CUBIC \ac{CE} gets generated. In addition, ROCCET ignores loss during Slow Start, since we observed performance issues in this phase. These issues come from a loss not generated by congestion or falsely detected by a retransmission timeout. One of the reasons for this can be an inter-arrival time jitter caused by the 5G L2 DRX mechanism or L2 HARQ retransmissions at the start of the transmission.\\

\noindent \textbf{\ac{ORBITER}} The congestion avoidance phase is entered after \ac{LAUNCH} using the window growth function of CUBIC to increase the \ac{cwnd}.
If the amount of incoming \acp{ACK} over 5 RTTs deviates more than 20\,\% from the \textit{cum\_cwnd} over the same time period, the first condition for a ROCCET \ac{CE} is fulfilled.
For the second condition, the relative change of the RTT must be higher than 100\,\%. In that case, a ROCCET \ac{CE} is generated. This \ac{CE} leads to a \ac{cwnd} reduction. To create a robust mechanism for cellular networks, the jitter of the connection is also taken into account for the srRTT bound of 100\,\%.
To drain the already bloated buffer, 100\,ms after a ROCCET \ac{CE} occurs, \ac{ROCCET} does not increase or decrease the \ac{cwnd}. 
On these \acp{CE}, CUBIC's $W_{max}$ is only set if the current \ac{cwnd} is larger than the old $W_{max}$. This results in a faster growth of the \ac{cwnd} after an ROCCET \ac{CE} occurs. 
On a ROCCET \ac{CE}, ROCCET reduces the \ac{cwnd} by the factor $\beta$. In addition, if a loss is detected, ROCCET behaves like CUBIC, reducing the \ac{cwnd} by the factor $\beta$ and doing fast convergence, if enabled.

\section{Real World Measurement Setup}
\label{sec:measurement_setup}
\begin{figure}
	\centerline{
		\resizebox{.85\linewidth}{!}{
			\includegraphics{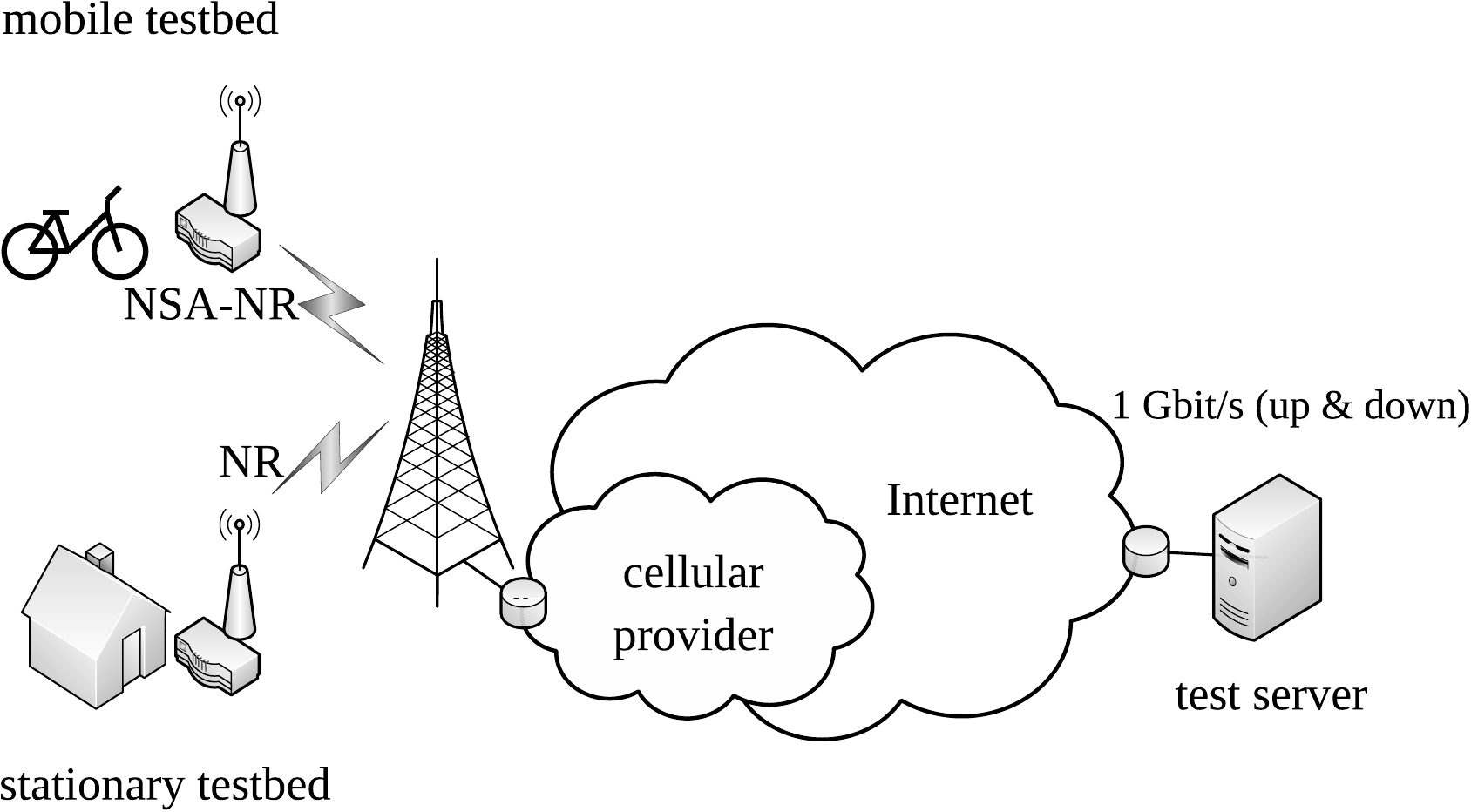}
		}
	}
	\caption{Measurement topology for mobile (standalone) and stationary downlink (non-standalone) measurements in two commercial \ac{NR}networks.}
	\label{fig:measurement-topo}
\end{figure}

\noindent In this section, we present our real-world 5G testbed. For evaluating the performance of \ac{ROCCET}, we used a \textit{Sierra Wireless EM9293} modem~\cite{em9293}. According to its specifications, the EM9293 can achieve download speeds of up to 4.9\,Gbps, and upload speeds of up to 1.3\,Gbps. The \ac{UE} was connected to a \textit{Raspberry Pi 5} running \textit{Debian GNU/Linux 12 (bookworm)} via a USB adapter.
For each measurement with \textit{iPerf3} \cite{iperf}, we have collected data on \ac{sRTT}, the send window, and the \ac{cwnd} size from the Linux debug module on the sender's side. On the receiver’s end, we used \textit{tcpdump} to capture \ac{TCP} traffic and documented the assigned/used \ac{NR} and \ac{LTE} component carriers, along with their bandwidths (expressed in MHz). Our two measurement campaigns were conducted within a major commercial \ac{NSA}-NR and \ac{SA}-NR network in Hannover, Germany. Due to the limited availability of these networks, we conducted our stationary measurements in an SA-NR and our mobile measurements in an NSA-NR network.  
The measurement setup is illustrated in Fig.~\ref{fig:measurement-topo}.
For mobile measurements, we selected a bike route that passes through a densely populated urban area with numerous buildings, as well as a large park. Our stationary \ac{UE} was placed on the 5th floor in an office building.
For evaluation, we compare \ac{sRTT} and achieved goodput from ROCCET, CUBIC, and BBRv3. We used BBRv3 as the third comparator, as BBRv1 is one of the most popular TCP congestion control algorithms, and BBRv3 is the latest version of \ac{BBR} currently in deployment.

\section{Stationary Measurements Results}
\label{sec:stat_measurements}
\begin{figure}[t!]
	\centerline{
		\resizebox{\linewidth}{!}{
			\includegraphics{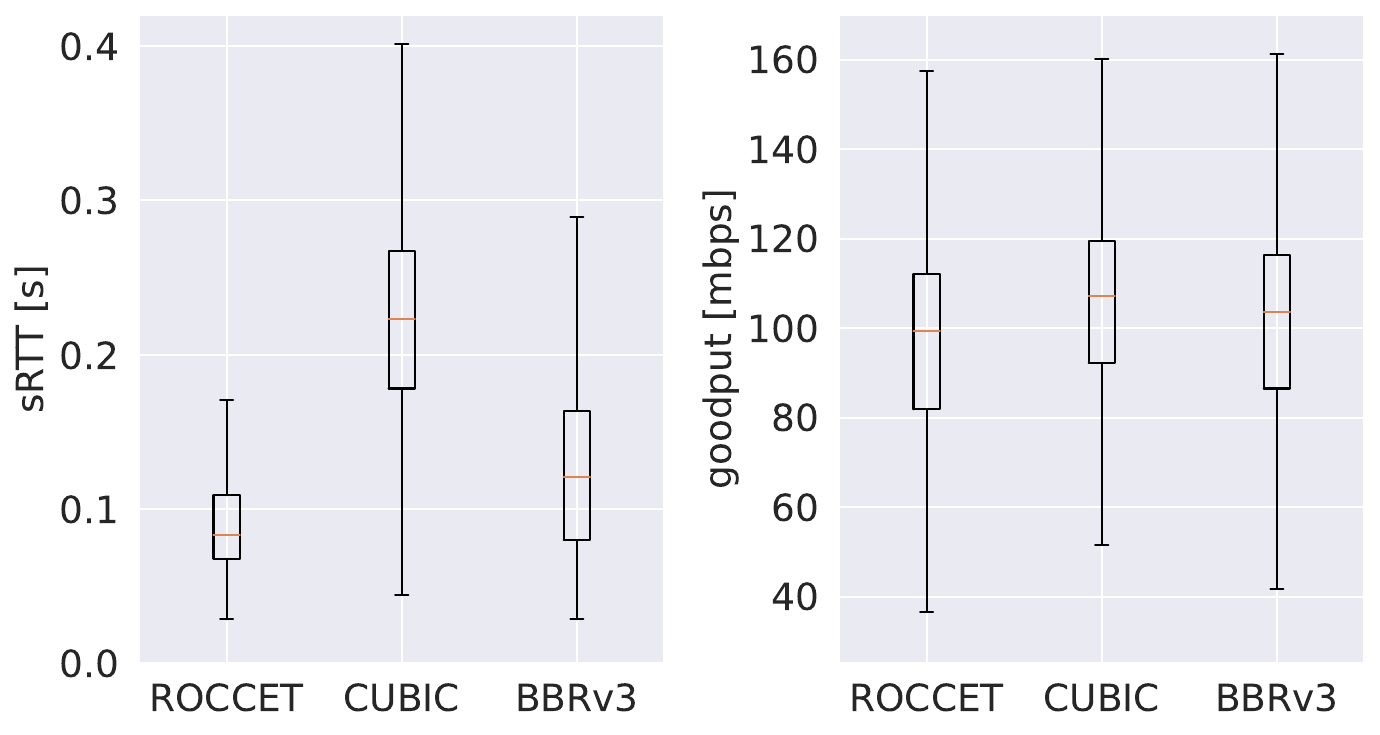}
		}
	}
	\caption{Goodput and sRTT boxplots for stationary 5G standalone greedy TCP throughput measurements in the downlink direction. Each measurement is 30\,s long; during the run, the congestion control was alternated. We did 120 measurements for each congestion control.}
	\label{fig:stationary-measurement-results}
\end{figure}

\noindent To evaluate the performance of ROCCET, we have performed 360 greedy throughput measurements in a commercial \ac{NR} standalone network. During the measurements, we have alternated between CUBIC, ROCCET, and BBRv3 (120 measurements for each).
Each measurement is 30\,s long, and since we want to improve the Slow Start performance with CUBIC, HyStart is disabled. Our results are shown in Fig. \ref{fig:stationary-measurement-results}.
When we compare ROCCET and CUBIC, it is observable that all \acp{sRTT} caused by ROCCET are below the 25\,\% quantile of CUBIC. A simple explanation for this is that for most transmissions with CUBIC, the \ac{cwnd} was frozen as we described in Sec \ref{sec:cc-issues}. Therefore, CUBIC was sending data at a steady rate and bloating the RLC buffer, similar to our example in Fig.~\ref{fig:cubic_slow-start_example}, Sec.~\ref{sec:cc-issues}. This is also the explanation for higher goodput with CUBIC, since the RLC buffer is never empty. In other words, CUBIC overfills the pipe and gets every available goodput, which is not detectable by ROCCET. Therefore, ROCCET's median goodput is nearly 10\,Mbps lower than CUBIC's. Nevertheless, ROCCET has a better performance than CUBIC, since ROCCET reacts earlier to changes in the network, like congestion, and causes less bufferbloat than CUBIC.

When we compare ROCCET with BBRv3, we observe that both variants can reach nearly the same goodput with a difference of 5\,Mbps in median. Also, the 25 and 75\,\% quantiles for BBRv3 are slightly higher. But when it comes to \ac{sRTT} 75\,\% of ROCCET's \acp{sRTT} are below the median of BBRv3. In addition, the maximum of ROCCET is slightly higher than the 75\,\% quantile for BBRv3. This can be explained through the variance of latency in cellular networks. Changes in latency are not only correlated with bottleneck link capacity. Therefore, BBRv3 causes some bufferbloat (but less than CUBIC), which leads to higher data rates and latency.


\section{Mobile Measurements Results}
\label{sec:mob_measurements}
\begin{figure}[t!]
	\centerline{
		\resizebox{\linewidth}{!}{
			\includegraphics{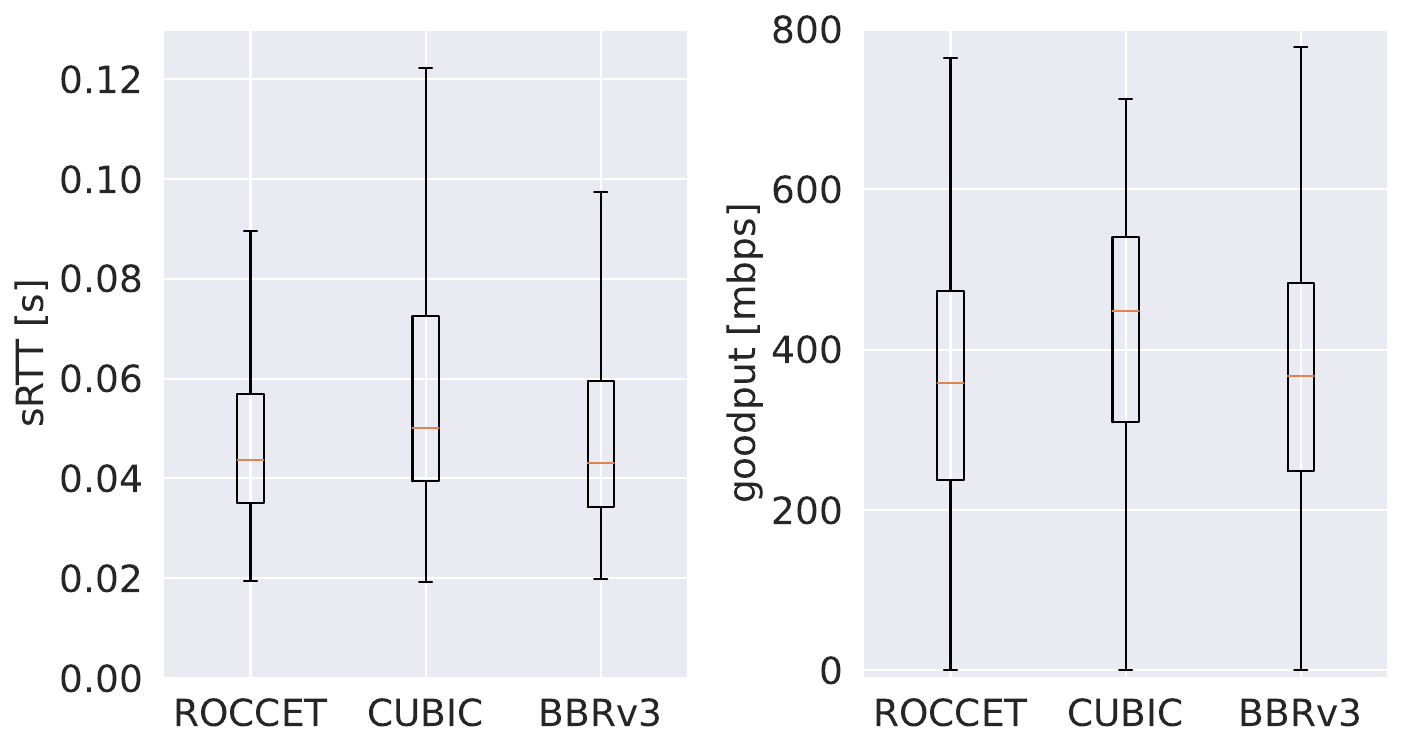}
		}
	}
	\caption{Goodput and sRTT boxplots for mobile greedy TCP throughput downlink measurements, where at least 90\,MHz of carrier bandwidth is available. Each measurement is 20\,s long, during the run, the congestion control was alternated. The measurement tour was more than 95\,min long, with at least 95 samples per congestion control.}
	\label{fig:mobile-measurement-results}
\end{figure}

\noindent Since we built ROCCET with the main intention of being used in cellular networks, we conduct mobile measurements in a commercial \ac{NR}-\ac{NSA} network. As for stationary measurements, we measured greedy TCP throughput. During the measurement ride, we performed 285 measurements. Each measurement is 20\,s long, and during each run, the congestion control was alternated. The results are filtered so that at least 90\,MHz of carrier bandwidth is available during transmission. We did this for a better comparison because data rates differ a lot if no 5G carrier is available during a measurement. The reached goodput and the respective sRTTs are shown in Fig.~\ref{fig:mobile-measurement-results}.
First of all, when we compare ROCCET with CUBIC, we can see that we have improved the sRTT. The median and 75\,\% quantile of \acp{sRTT} caused by ROCCET are lower than the 75\,\% quantile for CUBIC. In addition, the median of ROCCET is also lower than for CUBIC. As for our stationary measurements, the goodput with CUBIC is higher than with ROCCET. The reason for that is the same as in the previous Sec.~\ref{sec:mob_measurements}, CUBIC is overfilling the pipe, causing queuing.
Another result is that ROCCET and BBRv3 nearly have the same goodput distribution, with the exception that ROCCET causes slightly less latency than BBRv3. The 75\,\% quantile and maximum of ROCCET are lower than for BBRv3. On the other hand, the 25\,\% quantile and median for both are nearly the same. 

Given these points, we significantly improved CUBIC with our extension ROCCET. In addition, ROCCET has a performance that is at least as good as that of BBRv3.    

\section{TCP Bandwidth Share}
\label{sec:fairness}
\begin{figure*}

	\centerline{
        \subfloat[$\textit{50\,Mbps}\times\textit{30\,ms}$]{\label{fig:fairness-30ms-50Mbps}
    		
    			\includegraphics[width=9.3cm]{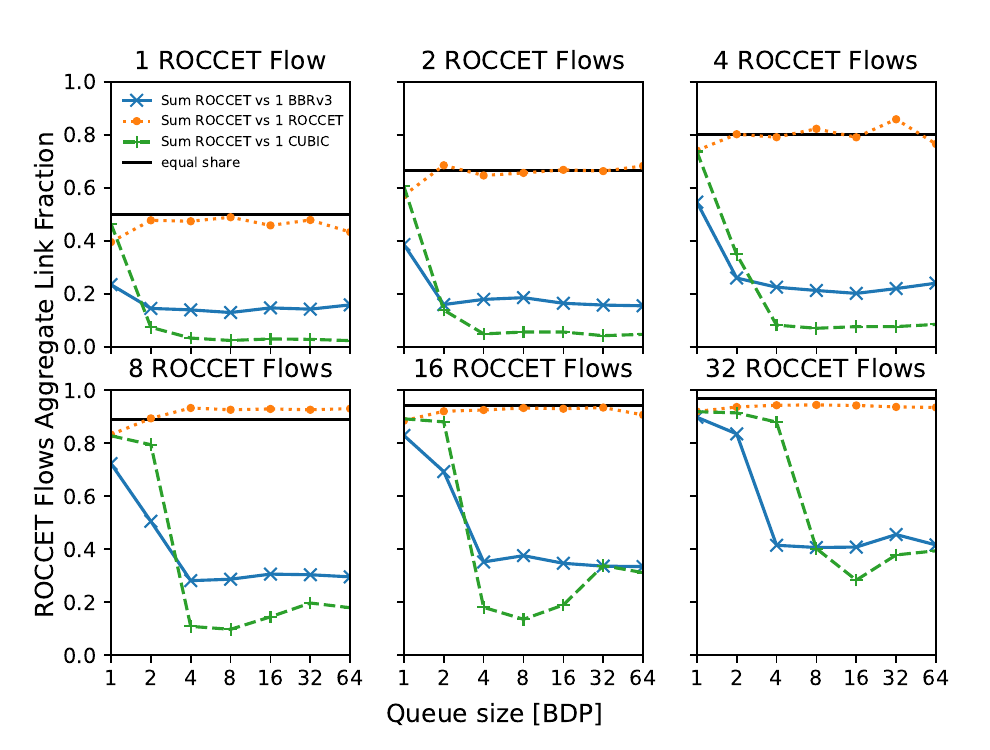}
    		
    	}
        \subfloat[$\textit{10\,Mbps}\times\textit{40\,ms}$]{\label{fig:fairness-40ms-10Mbps}
    		
    			\includegraphics[width=9.3cm]{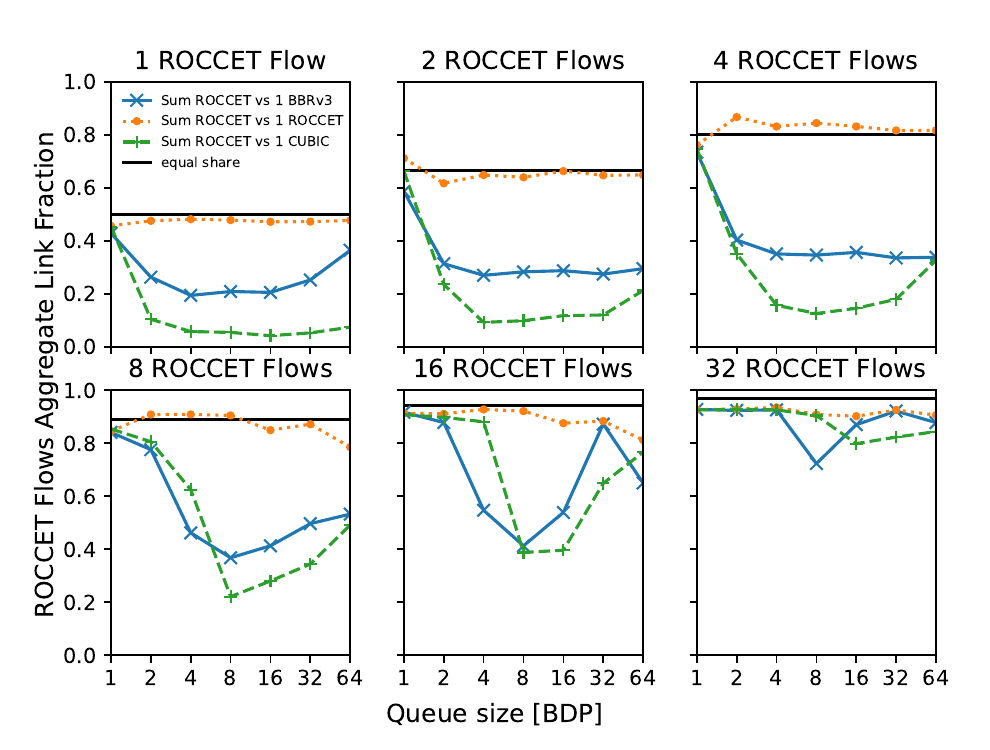}
    		
        }
    
}
	\caption{Bandwidth share of 1 to 32 ROCCET flows competing with a single BBRv3, CUBIC, or ROCCET flow. Link characteristics are $\textit{10\,Mbps}\times\textit{40\,ms}$ and~$\textit{50\,Mbps}\times\textit{30\,ms}$ with a bottleneck link buffer size of 1 to 32 BDP.}
	\label{fig:fairness}
\end{figure*}
\noindent In this section, we evaluate the bandwidth share of \ac{TCP} \ac{ROCCET} in competition with \ac{BBR}v3, CUBIC, and ROCCET itself. First, we evaluate the bandwidth share between multiple ROCCET flows competing with single BBRv3 and CUBIC flows. Afterwards, we analyse the inter- and intra-bandwidth share of single BBRv3 and CUBIC flows competing with \ac{ROCCET}.
We conducted these experiments based on the work of Ware et al.~\cite{r.ware-beyond-jain-fairness-index}.
With this evaluation, we want to show that ROCCET counts as a deployable congestion control algorithm.
Therefore, we conducted our measurements with the link characteristics of $\textit{50\,Mbps}\times\textit{30\,ms}$ and $\textit{10\,Mbps}\times\textit{40\,ms}$, using an Emulab \cite{emulab} link emulation.
We evaluated the bandwidth share at a bottleneck shared link in a dumbbell topology.

\subsection{Multi-Flow bandwidth share}
\label{subsec:fair-multi-flow}
\noindent For both link scenarios, we measure one competing TCP flow against 1 to 32 ROCCET flows, with a bottleneck link buffer size from 1 to 64 BDP. Since the buffers are deep, we set the receiver's receive window to a larger value than the buffer size. For each configuration, we did five measurements with a duration of two minutes. The results are shown in Fig. \ref{fig:fairness}. Each line represents the summed throughput of 1 to 32 flows from the same sender, which compete with a single BBRv3, CUBIC, or ROCCET flow from a different sender.
First of all can be seen that ROCCET has nearly an equal throughput share with other ROCCET flows, for both of our scenarios.

For the scenario $\textit{50\,Mbps}\times\textit{30\,ms}$: BBRv3 and CUBIC are getting more bandwidth than the equal share with \ac{ROCCET}. The bandwidth share with ROCCET is getting worse with the increase in the buffer size. Additionally,  CUBIC also gets more bandwidth than BBRv3 when competing with ROCCET flows. One explanation for this is that ROCCET is reacting to the bandwidth probing phase of BBRv3, detecting increasing RTTs and reducing the \ac{cwnd}. Such behavior is typical for congestion control algorithms that observe changes in RTTs like TCP Vegas \cite{vegas}. For CUBIC, the share is worse since CUBIC always tries to fill the buffer until a loss occurs. In this case, ROCCET reacts to queuing caused by CUBIC. If the buffer size increases further, the equal share with CUBIC gets better for 8 to 32 competing ROCCET flows.

For the scenario $\textit{10\,Mbps}\times\textit{40\,ms}$: Nearly the same behavior can be observed as for the previous scenario, with the addition of a better equal bandwidth share with BBRv3 and CUBIC. Also, the bandwidth share is increasing more with the buffer size than for the previous scenario. The conclusion from this is that for a smaller BDP and a larger buffer, the equal share increases.

Nevertheless, our results show that ROCCET has nearly an equal bandwidth share with other ROCCET flows for different link characteristics and small to large buffers.
On the other hand, BBRv3 and CUBIC obtain a larger bandwidth share than ROCCET when the buffer size exceeds one BDP. 
This effect increases with the size of the buffer. For cellular networks like 5G, this is not a problem since the network is scheduled; therefore, the bandwidth share depends on the implementation of the scheduler.
Furthermore, depending on the implementation of the \ac{RLC} layer, it is not even necessary that TCP flows for different receivers share the same \ac{RLC} layer buffer. In addition, it is also possible that on higher layers of the \ac{RAN} buffers are per single \ac{TCP} flow. An example of such an implementation can be found in OpenAirInterface~\cite{oai}. 

\subsection{Inter- and intra-bandwidth share}
\begin{figure}
\centerline{
        \subfloat[$\textit{50\,Mbps}\times\textit{30\,ms}$]{\label{fig:tcp-harm_30ms_50mbps}\includegraphics[width=.5\linewidth]{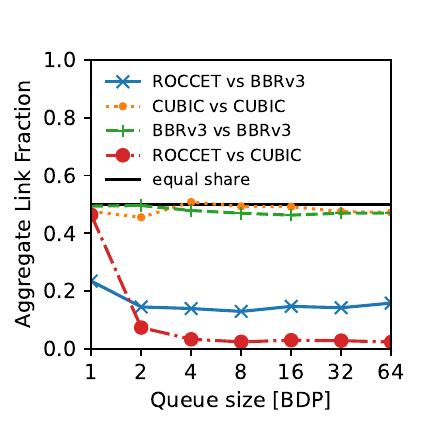}}
        
        \subfloat[$\textit{10\,Mbps}\times\textit{40\,ms}$]{\label{fig:tcp-harm_40ms_10mbps}\includegraphics[width=.5\linewidth]{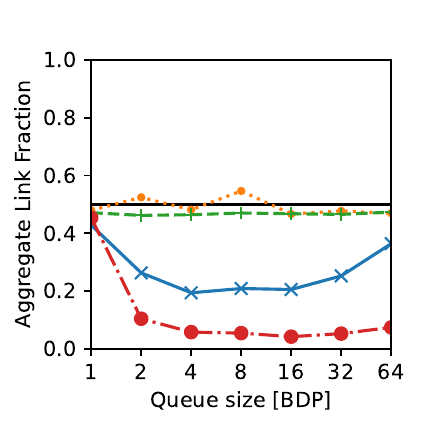}}
}
    \caption{Single flow inter- and intra-bandwidth share for BBRv3 and CUBIC with ROCCET for link charakeristics of $\textit{50\,Mbps}\times\textit{30\,ms}$ and $\textit{10\,Mbps}\times\textit{40\,ms}$.}
    \label{fig:harm}
\end{figure}

\noindent In addition to these bandwidth share experiments we have analyzed in Sec.~\ref{subsec:fair-multi-flow}, we conducted detailed analyses of the inter- and intra-bandwidth share behavior for single flows.
To this end, we measured the bandwidth share for CUBIC with CUBIC and BBRv3 with BBRv3 and compared the share with a single competing ROCCET flow. We evaluated these measurements for the same link characteristics as before, with a buffer size of 1 to 64 BDP. The bandwidth shares are shown in Fig.~\ref{fig:harm}.
Overall, it can be seen that BBRv3 and CUBIC provide a nearly equal share when competing with equal congestion control.
For a link of $\textit{50\,Mbps}\times\textit{30\,ms}$, the equal share of CUBIC with itself is nearly the same as for CUBIC and ROCCET if the buffer has a size of one BDP. Otherwise, ROCCET gets pushed out of the buffer by CUBIC, and the bandwidth share gets significantly worse.
For BBRv3, the bandwidth share of ROCCET is always less than 30\,\% of the link capacity, decreasing with increasing buffer size, but the bandwidth share between ROCCET and BBRv3 is always better than for ROCCET and CUBIC for buffer sizes above one BDP.
In addition, the decrease in bandwidth share for one and two BPD-sized buffers is less with BBRv3 than with CUBIC.

For the link characteristic of $\textit{10\,Mbps}\times\textit{40\,ms}$ we observe a better sharing behavior.
First of all, BBRv3 and CUBIC have the same bandwidth share with ROCCET as with each other at a buffer size of one BDP. As for our other measurements, the bandwidth share decreases with the increase in the buffer size. The behavior changes at a buffer size of 16 BDP. There, we observe an increase in the bandwidth share with increasing the buffer size.

Our results show that ROCCET is a defensive congestion control algorithm that allocates less than an equal bandwidth share if competing with CUBIC or BBRv3 flows in buffers larger than one BDP.
On the other hand, ROCCET can provide an equal or less intra-bandwidth share. These points lead us to conclude that ROCCET can be considered as a deployable congestion control algorithm since it causes less harm to other congestion control algorithms than other congestion control to themselves~\cite {r.ware-beyond-jain-fairness-index}. 

\section{Reaction to Bandwidth Changes}
\label{sec:bandwidth_change_reaction}
\begin{figure}

        \subfloat[CUBIC]{\label{fig:reaction-time-cubic}\includegraphics[width=\linewidth]{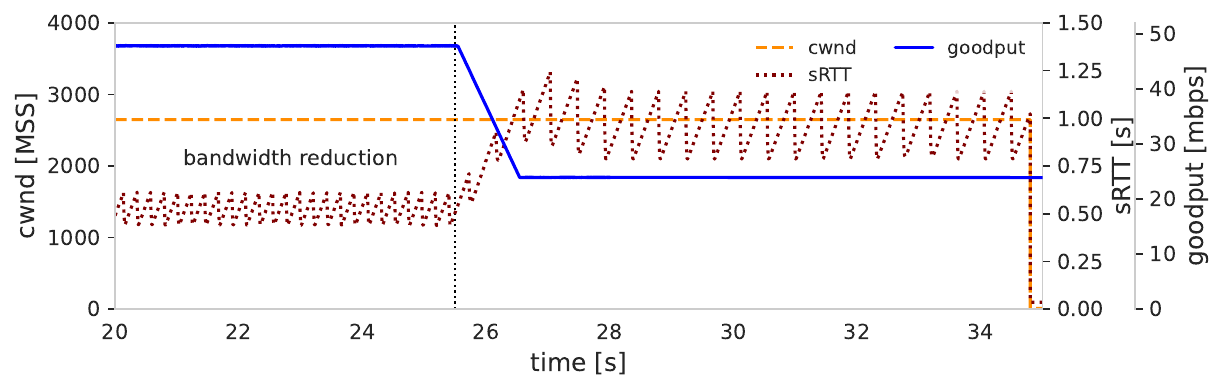}}
        
        \subfloat[ROCCET]{\label{fig:reaction-time-roccet}\includegraphics[width=\linewidth]{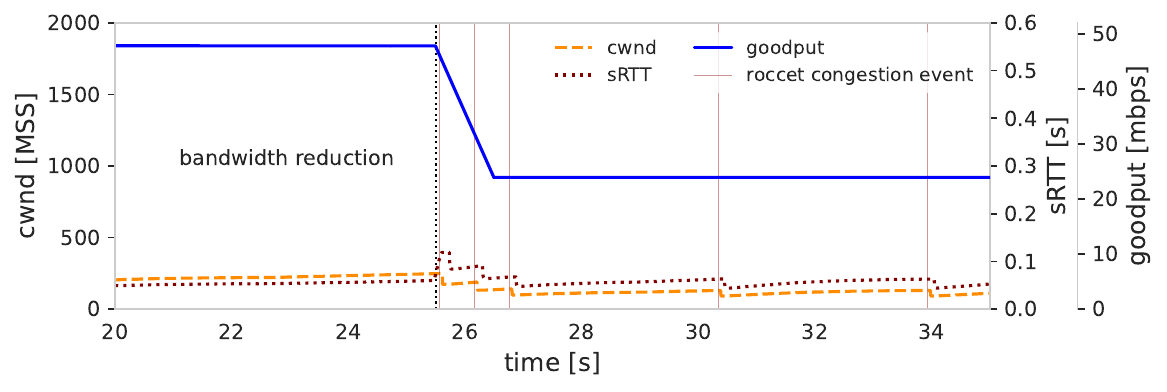}}

        \subfloat[BBRv3]{\label{fig:reaction-time-bbr}\includegraphics[width=\linewidth]{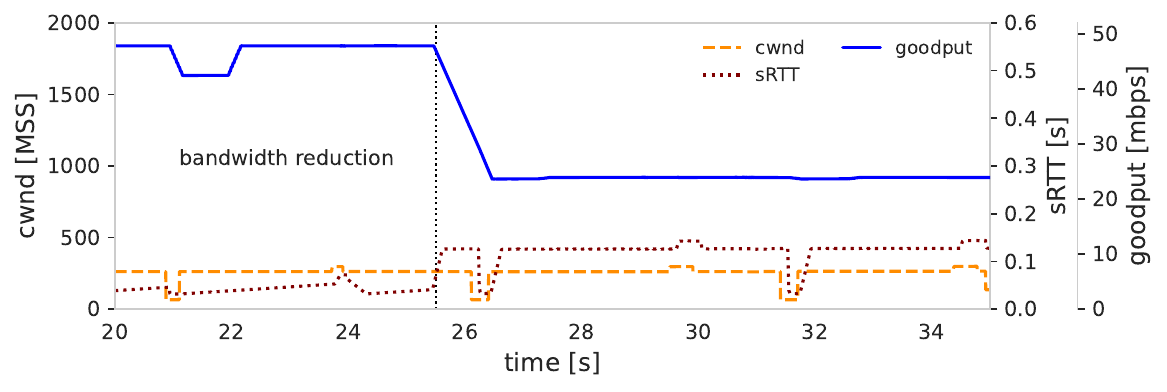}}

    \caption{Characteristic reaction of BBRv3, ROCCET, and CUBIC to bandwidth reduction from 50\,Mbps to 25\,Mbps with a \ac{RTT} of 40\,ms and a queue size of multiple BDP. Note: Different axis scaling for \ac{cwnd} and \ac{sRTT} in the transmission plot for CUBIC.}
    \label{fig:raction-time}
\end{figure}

\noindent Our last in-depth analysis of ROCCET compared to BBRv3 and CUBIC is the reaction to bandwidth reductions.
For this, we have performed greedy TCP throughput measurements for 35\,s with a link capacity reduction to half of the original link capacity. In our scenario, we choose a link characteristic of $\textit{50\,Mbps}\times\textit{30\,ms}$ with an oversized buffer of multiple BDP, which are usually found in cellular networks. The characteristic transmissions for BBRv3, CUBIC, and ROCCET are shown in Fig.~\ref{fig:raction-time}. For CUBIC, we have doubled the scaling of the axis for \ac{cwnd} and nearly tripled the axis for the \ac{sRTT}.
First of all, for CUBIC, we observe the unwanted side effect of the Linux kernel implementation, which we described in Sec.~\ref{sec:cc-issues},i.e., CUBIC does not react to changes in bandwidth. Instead, the goodput gets halved and the \ac{sRTT} increases.
This behavior gets solved by ROCCET. As we can see in Fig.~\ref{fig:reaction-time-roccet}, ROCCET generates several \acp{CE} after the bandwidth of the link is halved. In addition, ROCCET also generates \acp{CE} after the bandwidth reduction while bandwidth probing around 30 and 34\,s of the transmission.
When we compare the BBRv3 transmission (Fig.~\ref{fig:reaction-time-bbr}) with ROCCET (Fig.~\ref{fig:reaction-time-roccet}), we observe that both algorithms provide a nearly equal-sized \ac{cwnd} before the bandwidth reduction. After the bandwidth reduction, this behavior changes. \ac{BBR}v3 does not react to the halved bandwidth, which leads to higher \acp{sRTT} compared to ROCCET. Furthermore, BBRv3 does not detect the bandwidth reduction until the end of the transmission, even though it enters the min RTT probing state for two times. Some similar performance issues were also observed by \cite{Promises-and-Potential-of-BBRv3}. If the available bandwidth gets reduced, BBRv3 has substantial delays in \ac{cwnd} adoptions.

\section{Implementation}
\label{sec:implementation}
\noindent In this section, we briefly describe the implementation of \ac{TCP} ROCCET and all additional software we provide for development and debugging the kernel module. Our implementation can be found in a Git repository \cite{roccet-git}. Since we have implemented ROCCET as a pluggable kernel module, there is no need to build a new kernel. This means that TCP ROCCET can be loaded while the operating system is running. 
Our implementation generally extends the Linux kernel code for \ac{TCP} CUBIC. Most of our code was added to the \ac{TCP} \textit{.cong\_avoid} function. In addition, we added functions to update internal parameters such as $RTT_{min}$ and $ack\_rate$.
Our implementation contains three files: \textit{tcp\_roccet.c}, containing the extension of CUBIC, \textit{tcp\_roccet.h}, which contains the internal struct needed for the algorithm, and \textit{roccet\_kprobe.c}, an additional kernel module to provide live data of a ROCCET socket during a transmission. The ROCCET kprobe module works similarly to the already existing in the kernel \ac{TCP} debugging tools.

\section{Paramteter Options}
\label{sec:discussion}
\noindent Our results were generated using the ROCCET algorithm described in Sec.~\ref{sec:roccet-algo}.
Our implementation offers the possibility to adjust several of ROCCET's parameters.
Making the algorithm less aggressive by detecting an increase in sRTT, leading to less \acp{CE} generated by ROCCET. The two parameters are the tolerance of the received ACK count and the upper bound for the srRTT, at which an ROCCET \ac{CE} is generated. By increasing these parameters, ROCCET gets more defensive if the sRTT increases.
Furthermore, there is the possibility to ignore loss as a \ac{CE} generally, and, in addition, a recalculation of the $RTT_{min}$.
The recalculation works as follows: If the last update of $RTT_{min}$ is more than ten seconds ago, \ac{ROCCET} calculates a new $RTT_{min}$ by using an \ac{EWMA}, with $RTT_{min}$ and the current RTT as parameters.
For a scenario where RTTs increase slightly over a longer time interval of several seconds, this recalculation of the $RTT_{min}$ causes a steady state of srRTT. This can lead to poorer detection of increasing RTTs.
Nevertheless, we provide these configuration options to adjust the behavior of ROCCET if they are needed for special link characteristics, for example, lossy links.
Since ROCCET is an extension of CUBIC, all inherited CUBIC parameters can be adjusted.



\section{Conclusion}
\label{sec:conclusion}
\noindent In this paper, we presented a new TCP congestion control algorithm suited for current cellular 5G \ac{NR} networks. It extends the Linux kernel default congestion control CUBIC and improves its performance, and additionally solves the unwanted side effects of CUBIC's implementation, which we also discuss in this paper.
We have described our algorithm in detail and demonstrated its performance through both stationary and mobile measurements in two different commercial networks.
Taking our measurement results into account, ROCCET significantly improves the performance of CUBIC by reducing the latency. 
In addition, we have shown that ROCCET is able to achieve the same goodput as BBRv3 but with less latency.
Furthermore, we have performed bandwidth share experiments comparing ROCCET with CUBIC and BBRv3. We provide an in-depth analysis of multi- and single-flow bandwidth share, respecting the work from Ware et al. \cite{r.ware-beyond-jain-fairness-index}, which consequently classifies our congestion control as deployable, 
since ROCCET's inter-bandwidth share with BBRv3 and CUBIC is at least as good as the intra-bandwidth share of BBRv3 and CUBIC. 
Afterwards, we have analyzed the reaction of TCP ROCCET to bandwidth changes compared to BBRv3 and CUBIC. Our analysis shows that TCP ROCCET provides a better performance than the compared congestion control.
Furthermore, our kernel module offers several configuration options that can be changed.
For future work, we will take a look at ROCCET's performance on different kinds of networks and further optimize its behavior.
Our implementation of ROCCET can be found here \cite{roccet-git} as a kernel module. As a long-term goal, we will add the code to the official Linux repository. Additionally, a debug module is available in our repository, providing insight into ROCCET's internal variables, such as srRTT.

\bibliographystyle{IEEEtran}
\bibliography{bibliography}

\end{document}

%% file: figure/roccet_flow_chart.tex
\begin{tikzpicture}[
    auto,
    node distance = 11mm and 22mm,
      base/.style = {draw, rounded corners=5pt,
                     align=center},
     block/.style = {base, minimum width=3.5cm},
smallblock/.style = {base, minimum width=3.5cm},
                        ]
\coordinate (in);
        \node[block] (update) {update();};
        \node[above=5mm of update] (start) {on ACK received};
        \node[block, below=5mm of update] (is-in-slowstart) {is in slow start ?};
        \node[block, below=5mm of is-in-slowstart] (ss-srrtt) {$\textit{srRTT} >= 100\,\%$};
        \node[block, below=5mm of ss-srrtt] (ss-ackrate) {$\textit{recv ACKs} - cum\_cwnd$ \\ $>= 10\,\textit{segments}$};
        \node[block, below=5mm of ss-ackrate] (ss-init) {is initial slow start ?};
        \node[block, fill=lightgray, below=5mm of ss-init] (half-cwnd) {$cwnd = \frac{cwnd}{2}$ \\ $ss\_thresh = cwnd$};

        \node[block, right=13mm of is-in-slowstart] (drain) {is in drain ?};
        \node[block, below=5mm of drain] (5rtt_ago) {is last check 5 RTT ago ?};
        
        \node[block, below=5mm of 5rtt_ago] (roc-srrtt) {\textit{srRTT} $>= 100\,\%$};
        \node[block, below=5mm of roc-srrtt] (roc-ackrate) {$cum\_cwnd - \textit{recv ACKs}$ \\ $> 20\,\% \cdot cum\_cwnd$};
        \node[block, fill=lightgray, below=5mm of roc-ackrate] (roc-ce) {ROCCET CE};
         \node[block, fill=lightgray, below=5mm of roc-ce] (cubic-ce) {CUBIC CE};

         \node[block, fill=lightgray, below=5mm of cubic-ce] (cubic) {CUBIC congestion control};

          \node[block, fill=lightgray, above=5mm of drain] (return) {return};




        \draw [-{Latex[length=3mm]}] 
            (update) edge (is-in-slowstart) 
            (is-in-slowstart) edge ["Yes"] (ss-srrtt)  
            (ss-srrtt) edge ["Yes"] (ss-ackrate) 
            (ss-ackrate) edge ["Yes"] (ss-init)
            (ss-init) edge ["Yes"] (half-cwnd)

            (is-in-slowstart) edge ["No" above] (drain)
            (drain) edge ["No"] (5rtt_ago)
            (5rtt_ago) edge ["Yes"] (roc-srrtt)
            (roc-srrtt) edge ["Yes"] (roc-ackrate)
            (roc-ackrate) edge ["Yes"] (roc-ce)
            (drain) edge ["Yes"] (return);

        \draw [-{Latex[length=3mm]}] 
        (ss-init.west) -- +(-.5,0) |- node[below,pos=0.72] {No} (cubic-ce.west);

        \draw [-{Latex[length=3mm]}] 
        (5rtt_ago.west) -- +(-.6,0) |- node[pos=0.25] {} (cubic.west)
        (roc-srrtt.west) -- +(-.84,0) |- node[pos=0.2] {} (cubic.west)
        (roc-ackrate.west) -- +(-.69,0) |- node[pos=0.25] {} (cubic.west);

        \draw [-{Latex[length=3mm]}]
        (start.east) -- +(1,0) |- node[pos=0.25] {} (update);

        \draw [-{Latex[length=3mm]}]
        (ss-srrtt.west) -- +(-0.6,0) |- node[below,pos=0.735] {No} (cubic.west);
        \draw [-{Latex[length=3mm]}]
        (ss-ackrate.west) -- +(-0.44,0) |- node[pos=0.5] {} (cubic.west);

        \draw[thick,dotted] ($(ss-srrtt.north west)+(-0.2,0.4)$)  rectangle ($(half-cwnd.south east)+(0.2,-0.8)$) node[below, xshift=-1.9cm] {\textbf{LAUNCH}};

        \draw[thick,dashed] ($(5rtt_ago.north west)+(-0.4,0.4)$)  rectangle ($(cubic.south east)+(0.2,-0.3)$) node[below, xshift=-2.1cm] {\textbf{ORBITER}};
       
    \end{tikzpicture}